\documentclass[fleqn,twoside]{article}
\usepackage{espcrc2}
\usepackage{epsfig,amssymb,amsmath}
\newcommand{\msb}{{\overline{MS}}}
\newcommand{\psib}{{\overline{\psi}}}

\title{ Progress Towards finding Quark Masses 
 and the QCD scale  $\Lambda$ from the Lattice}

 \author{ 
     P.E.L. Rakow\address{ %Theoretical Physics Division, 
 Department of Mathematical Sciences, University of Liverpool,
 Liverpool L69 3BX, UK} }
       
\begin{document}

\begin{abstract}
 We discuss recent work trying to extract the renormalised
 quark masses and  $\Lambda$, 
 the QCD scale, from dynamical simulations in lattice 
 gauge theory.  
\end{abstract}

\maketitle

\vspace*{-6.9cm}
 {\large LTH 644}
 \vspace*{5.5cm}

 \section{INTRODUCTION} 

   Quark masses and $\Lambda_{QCD}$ are both parameters which are
 not accessible to direct observation, unlike quantities such as
 hadron masses which have a very clear meaning both in the 
 continuum and on the lattice. Extraction of either quantity
 from lattice data inevitably involves both a matching of
 lattice measurements with continuum measurements
 and the use of results from perturbation theory. 

     We can divide the reliance on perturbation theory into 
 two classes. In the most favourable cases all the perturbation
 theory needed can be performed in the continuum. In this case 
 the technology of perturbation theory is very well-developed,
 and we can expect to find fairly long perturbative series
 (often to three loops) for the quantities we need.  Less 
 favourable is the case when lattice perturbation theory is needed. 
 The propagators and vertices for lattice perturbation theory 
 are much more complicated,  making high order calculations 
 very difficult. However we will see that some impressive
 progress is being made with lattice perturbation theory too.

    The structure of this paper is as follows. 
 First we will discuss renormalisation in general, as this is
 crucial to producing quark masses defined in a way that 
 can be related to continuum physics. 
 Then we will discuss quark masses, dealing separately with the
 lighter quarks (up, down and strange) and (in less detail)
 with the heavier quarks (charm and bottom). 

    For the light quarks the method we discuss in most detail is the 
 use of the pseudoscalar meson masses to determine the quark masses. 
 From the lowest-order chiral perturbation theory we know that 
 a pseudoscalar meson made of quarks of flavour $a$ and $b$ has
 its mass given by 
 \begin{equation}
 m_{PS}^2 \propto m_a + m_b 
 \label{chiral_mps} 
 \end{equation} 
 which implies that the pseudoscalar meson mass (and especially
 the $\pi$ mass) will be very sensitive to the quark masses. 
 Eq.~(\ref{chiral_mps}) also implies that the average light quark mass
 $m_l \equiv \frac{1}{2} (m_u + m_d)$ will be easier to measure
 than the separate $u$ and $d$ masses, and in this contribution 
 we will only discuss this averaged light quark mass. 

    In the second part of the paper we will discuss 
 attempts to fix the scale parameter of QCD, $\Lambda_{QCD}$, 
 both from measurements carried out at ``normal'' lattice spacings,
  using gauge configurations generated for other projects, and
 from fine lattice spacings, using simulations dedicated to the purpose of
 determining  $\Lambda$.

%----------------------------------------
   \section{RENORMALISATION AND $Z$ FACTORS} 

 With very few exceptions, an operator expectation value 
 measured on the lattice has to be renormalised before we
 can give a number which can be compared with experiment.  
 One can think of this as a calibration factor. 
 Estimating $Z_m$, the renormalisation factor for the quark mass,
 is one of the more difficult steps in the calculation of 
 quark masses, so it is appropriate to discuss renormalisation
 in some detail. 

    There are a handful of cases where the $Z$ factor is
 not needed (for example, the conserved vector current). 
 In other cases one can do a completely  non-perturbative 
 calculation. An example of this is $Z_V$, the renormalisation
 constant for the local vector current, 
  $\overline{\psi} \gamma_\mu \psi$. We know, from the conservation
 of  baryon number  and charge, exactly what 
 the correct answer should be when the matrix element of
 the vector current is measured in a hadron. This can be used
 to calculate $Z_V$, e.g.~\cite{QCDSF_ZV}. This calculation does not rely on
 any perturbation theory at all, neither on the lattice nor in the
 continuum. 
 
    Usually we are not this fortunate, and we have to calibrate 
 our lattice probe by comparing lattice results 
 with continuum  perturbation theory results~\cite{Martinelli:1994}. 
 To use this method one measures quark propagators and Greens
 functions at a range of virtualities. Because these are gauge-dependent
 quantities the gauge must be fixed, the usual choice is to 
 impose the Landau gauge. 
 The relation we use to define the renormalisation of an 
 operator $O$ is
 \begin{equation} 
 \Lambda_O^{\overline{MS}}
 = \frac{Z_O}{Z_\psi} \Lambda_O^{lat} , 
 \end{equation} 
 where $Z_\psi$ is the wave-function renormalisation constant, 
 defined in a similar fashion 
 by comparing the lattice propagator with the ${\overline{MS}}$
 propagator\footnote{Traditionally the process is split up into 
 two stages by introducing an intermediate renormalisation scheme
 such as RI or RI$^{\prime}$. A renormalisation factor taking us from
 lattice to  RI$^{\prime}$ is followed by a conversion factor 
 to change  RI$^{\prime}$ to  ${\overline{MS}}$. I think the
 nature of the calculation is easier to follow if we combine
 both stages, and convert directly from lattice to  ${\overline{MS}}$.}. 

    Where do we get the  ${\overline{MS}}$ Greens functions
 which we need for this calculation? Here we have to rely on 
 continuum perturbation theory. The results we need to
 interpret our lattice results are now available up to 
 3 or 4 loops~\cite{MSbarGreen}. 
  Even the continuum perturbation series can be slowly converging
 in some cases. Tricks to accelerate the series' convergence,
 such as careful choice of scale, may be helpful.

 The comparison between lattice and continuum results must
 be performed at a scale $\mu$ that satisfies
 \begin{equation} 
   \Lambda^2_{QCD} \ll \mu^2 \ll 1/a^2 .   
 \end{equation} 
 The lower limit arises because when $\mu^2$ is too small we do not
 know the true value of $\Lambda_O^{\overline{MS}}$, since our 
 perturbative series do not converge fast enough, and because the
 Greens function may have large non-perturbative contributions, 
 for example contributions associated with the spontaneous breaking 
 of chiral symmetry and the existence of a quark condensate. 
   The upper limit arises because the lattice Greens function 
 will have discretisation errors which become important when
 $ a^2 \mu ^2 \sim 1$. The method of~\cite{Martinelli:1994}
 relies on the existence of a plateau region between these two
 limits, and in practice this may turn out to be disappointingly
 narrow.   
   
    There is little that can be done about the lower limit, 
 what happens here is a real physical effect, present both on the
 lattice and in the continuum. On the other hand lattice perturbation
 theory can be used to greatly reduce the $ a^2 \mu ^2 $ errors, 
 extending the useful plateau region outwards.

\begin{figure}[htbp]
 \begin{center}
\includegraphics*[width=5cm,angle=270]{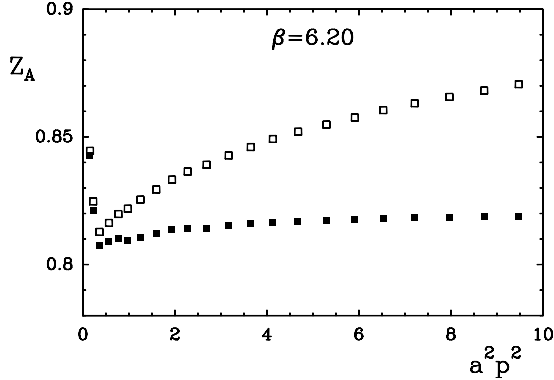}
 \vspace*{-4mm}
\caption{ \it The renormalisation factor for the axial current,
 for quenched clover fermions at $\beta=6.20$. The open squares
 show the raw data, the black squares show the result after 
 subtraction of the one-loop discretisation errors. The subtraction
 procedure increases the plateau region.\label{ZAsub} }
 \vspace*{-8mm}
 \end{center} \end{figure}

   Normally lattice perturbation theory results are quoted for small
 external momenta   $ a^2 \mu^2 \ll 1 $  but with a little extra
 effort the Greens functions can be found for any external momentum. 
 Subtracting off the one-loop contribution to the discretisation errors 
 allows us to reduce errors of this type from 
   $O(g^2 a^2 \mu^2)$ to $O(g^4 a^2 \mu^2)$. 
 The success of this procedure can be seen in Fig.~\ref{ZAsub}, 
 which shows a much better plateau after subtraction of one-loop
 lattice artefacts.

% Non-perturbative comparison of RGI and lattice results
% \begin{center}
%\includegraphics[width=8cm]{zs_540_1350.ps}
%\end{center}

%-------------------------------------------------

 \subsection{ Singlet Contribution } 

   We have discussed how to measure the renormalisation 
 constants for quark bilinear operators, but how can we measure
 the mass renormalisation factor $Z_m$? The usual method is to 
 look for an identity relating  $Z_m$ to $Z_S$, 
 the renormalisation factor for the scalar operator. 

   The usual way to derive this is by looking at the divergence
 of a flavour non-singlet vector current $J_\mu^a$. 
 In the continuum this current is $\psib \gamma_\mu \tau^a \psi$, 
 where $\tau$ is a traceless flavour matrix. 
 The lattice form depends on the fermion action, for Wilson
 or clover fermions it is\\
 $\frac{1}{2} \psib(x+\hat\mu) [\gamma_\mu +r]  \tau^a \psi(x)
 $\\
 $\hbox{}\qquad \qquad + \frac{1}{2}\psib(x)  [\gamma_\mu-r] U^\dagger(x)
 \tau^a \psi(x+\hat\mu) . $ \\
 The Ward identity~\cite{MM} reads
 \begin{equation}
 \left\langle \partial_\mu J_\mu^a \right\rangle
 = \left\langle \psib \;  \left[ M , \tau^a \right] \;
 \psi  \right\rangle
 \hbox{ + contact terms,} 
 \label{VWIeq}
 \end{equation}  
 where $M$ is the quark mass matrix. This identity links the flavour
 non-singlet scalar operator and the flavour non-singlet part of
 the mass matrix. Eq.~(\ref{VWIeq}) is an exact lattice
 identity if we use the conserved vector current, but because it 
 only involves the non-singlet part of the mass matrix (i.e.~mass
 differences), it still permits an additive renormalisation
 of the singlet part of the mass matrix (which does of course
 occur in Wilson or clover fermions). 

  Requiring that eq.~(\ref{VWIeq})
 still hold after renormalisation tells us that
 $ Z_m^{NS} Z_{\psib \psi}^{NS} = 1 $.  This tells us how to 
 renormalise non-singlet quantities, such as mass differences, 
 but it doesn't tell us how the singlet part of $M$ (which
 commutes with $\tau$) renormalises. 

\begin{figure}[htbp]
 \begin{center}
\includegraphics*[width=0.15\textwidth, angle=270]{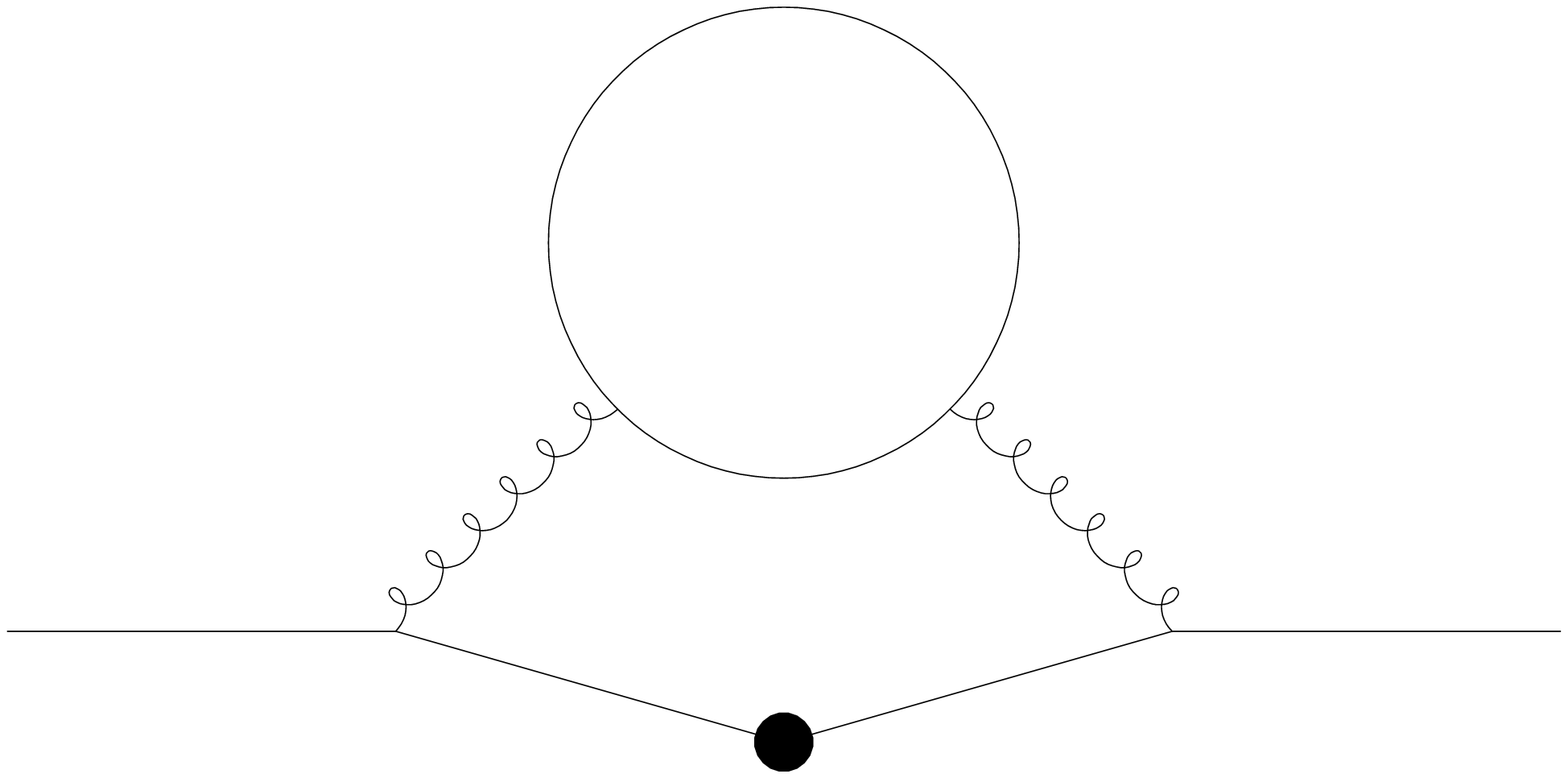}\bigskip\\
 % \hbox{ } \\
\includegraphics*[width=0.15\textwidth, angle=270]{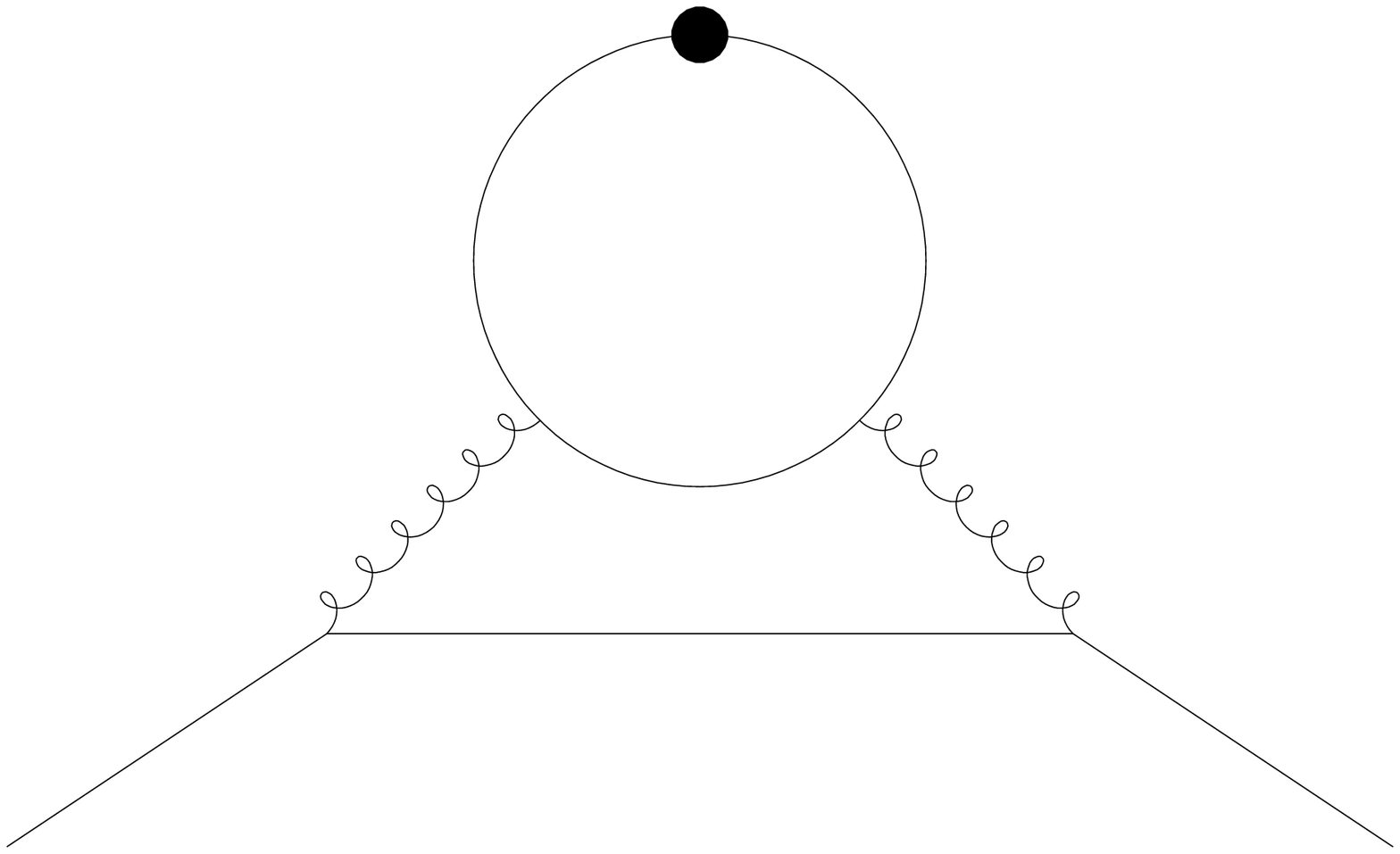}
 \end{center} 
 \vspace*{-9mm}
 \caption{ \it An example (above) of a Feynman diagram 
 present in both singlet and non-singlet cases, and
 (below) a diagram which only contributes to the singlet $Z$. 
 The black point shows where the $\psib \psi$ operator is inserted. 
 \label{discon_diag} } 
 \vspace*{-4mm}
 \end{figure}

 In $\overline{MS}$ we are used to having a difference
 between singlet and non-singlet renormalisation factors
 for operators with an {\it odd} number of gamma matrices. 
   Because clover fermions have no exact chiral symmetry, 
 there can be a difference between singlet and
 non-singlet $Z$ for operators with an {\it even}
 number of gamma matrices too. 
 This can be seen in Fig.~\ref{discon_diag}. In the continuum
 the ``bubble'' diagram would be the product of 5 gamma matrices
 (3 from propagators, 2 from $qqg$ vertices) so its trace would be
 zero in the massless case. For Wilson or clover fermions this
 no longer holds, and the bubble will give a non-zero contribution. 

   The singlet vector current does not give us any useful information
 about the mass matrix, but we can find some relationships
 by considering partially quenched QCD, with valence quarks allowed
 to have  a different mass from the sea quark mass. 
 If we change the valence quark mass while keeping the sea quark mass
 fixed, the derivative of the quark propagator $S$ gives the 
 non-singlet scalar three-point function
 \begin{equation} 
  \frac{\partial}{\partial{\frac{1}{2\kappa_{val}}}} S(p)
 = G^{NS}_{\psib \psi}(p).       
 \label{NSid} 
 \end{equation} 
 However if we change valence and sea quark masses together we 
 get an identity for the singlet scalar  three-point function
 \begin{equation} 
  \left( \frac{\partial}{\partial{\frac{1}{2\kappa_{val}}}} 
 +  \frac{\partial}{\partial{\frac{1}{2\kappa_{sea}}}} \right) S(p) 
 = G^{S}_{\psib \psi}(p) \; .
 \label{Sid}
 \end{equation} 
 From these identities we conclude that 
   $ Z_m^{NS} Z_{\psib \psi}^{NS} = 1 $ (which we already knew from 
 the Vector Ward Identity) and  $Z_m^{S} Z_{\psib \psi}^{S} = 1$ 
 which is new. We can find $Z_{\psib \psi}^{NS}$ and thus 
  $ Z_m^{NS}$ by the non-perturbative methods discussed earlier. 
  The identities here also give us a way of finding  $Z_m^{S}$. 

 \begin{figure}[htbp]
 \begin{center}
 \includegraphics*[width=5cm,angle=270]{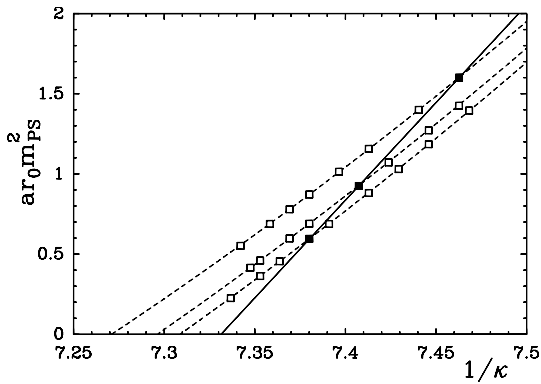}
 \caption{\it Dynamical and partially quenched pseudoscalar meson
 masses for clover fermions at $\beta=5.29$. Data from~\cite{QCDSFmq}. 
 \label{PQdata} } 
 \end{center}
 \vspace*{-6mm} 
 \end{figure}

 In Fig.~\ref{PQdata} we show partially quenched (open points) and
 full dynamical (filled points) data for the pseudoscalar meson 
 masses. We can immediately see that there is a different critical
 $\kappa$ if we follow the solid line connecting
  the full dynamical points (the points with
 valence and sea quark masses equal) than if we follow one of the 
 dashed lines (sea quark mass held fixed, valence quark mass varied). 
 This is because clover fermions have no exact chiral symmetry, 
 and so an additive quark mass renormalisation is possible. 
 The difference in critical $\kappa$s shows that this additive 
 term depends quite strongly on the sea quark mass. 

    In the continuum the situation would look quite different ---
 because of chiral symmetry both the full and partially quenched
 curves would both have to pass through the origin, at 
 zero valence quark mass. This would also apply for a lattice
 fermion formulation (such as overlap fermions) which have an
 exact chiral symmetry. After renormalisation the lattice 
 results ought to show the same structure, with both full dynamical
 and partially quenched $m_{PS}$ vanishing at the same place. 
 The only way to arrange this is to use different renormalisation
 factors for the partially quenched and full QCD quark masses,
 see Fig.~{\ref{FullPartial}, just as we expected from considering
 the identities eqs.~(\ref{NSid}) and (\ref{Sid}). 

 \begin{figure}[htbp]
 % \begin{center}
 \includegraphics*[width=0.33\textwidth,angle=270]{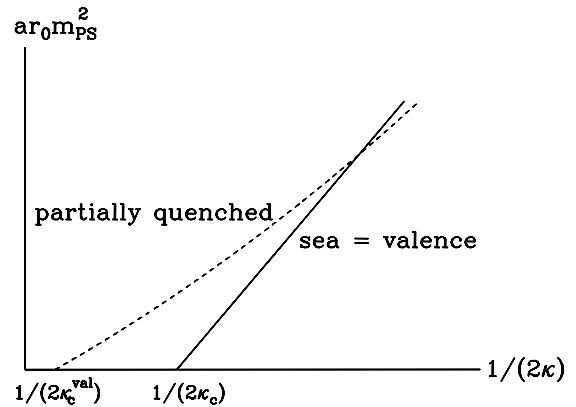}\\[2em]
 
 \includegraphics*[width=0.33\textwidth,angle=270]{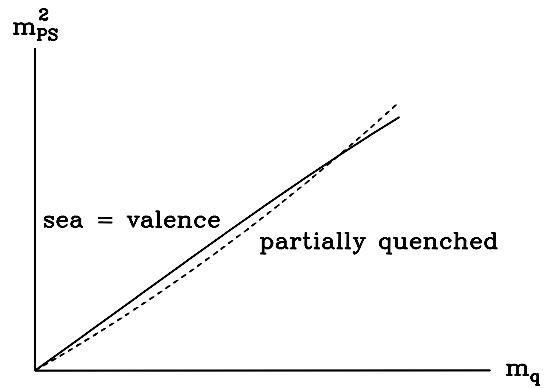}
 % \end{center}
 \caption{ \it Dynamical and partially  quenched pseudoscalar meson
 masses before and after renormalisation of the quark mass. 
  \label{FullPartial} } 
 \end{figure} 

     We define the bare sea and valence quark masses by
 \begin{equation}
  m = \frac{1}{2 \kappa} -  \frac{1}{2 \kappa_c},
 \end{equation} 
 with $\kappa_c$ defined as the critical $\kappa$ for QCD with
 equal valence and sea quark masses (see Fig.~{\ref{FullPartial}). 
 The renormalised masses are then defined by
 \begin{eqnarray}
 m_{sea}^R &=& Z_m^S m_{sea} \; ,\label{renrecipe}\\
 m_{val}^R &=& Z_m^{NS} ( m_{val} - m_{sea} )  +  Z_m^S m_{sea} . 
 \nonumber
 \end{eqnarray} 
 The partially quenched pseudoscalar mesons are massless at the
 point where $m_{val}^R =0$, which we call $\kappa_c^{val}$. 
 We can use this to find the ratio of the two $Z$'s, 
 \begin{eqnarray} 
 \lefteqn{\frac{  Z_m^S}{ Z_m^{NS}} 
  =  \left. \frac{ m_{sea} - m_{val} }{m_{sea}} 
  \right|_{\kappa_{val} = \kappa_{val}^c} } \\ && \nonumber \\
 &=& \left( \frac{1}{2 \kappa_{sea} } -  \frac{1}{2 \kappa_{val}^c }\right)
 \left( \frac{1}{2 \kappa_{sea} } -  \frac{1}{2 \kappa^c }\right)^{-1}
 \;. \nonumber
 \end{eqnarray}
 $Z$ ratios calculated this way in~\cite{QCDSFmq} are shown in 
 Fig.~\ref{rationew}.  The ratio only depends weakly on the 
 quark mass, but it depends very strongly on $\beta$. The ratio
 drops rapidly as lattice spacing decreases.    

 \vspace*{-0.5cm}
 \begin{figure}[htbp]
 \begin{center}
 \vspace*{-1.5cm}
 \includegraphics*[width=0.45\textwidth]{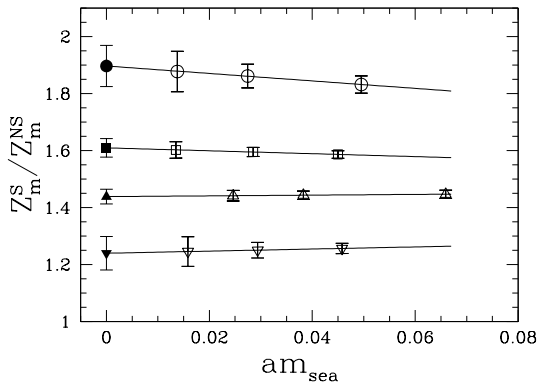}
 \caption{\it The ratio $Z^S_m/Z^{NS}_m$ for dynamical clover
 fermions~\cite{QCDSFmq}. 
 $\beta$ values run from 5.20 (highest line) to 5.40 (lowest line).
 \label{rationew} } 
 \end{center} 
 \end{figure}

    One way to illustrate the importance of the ratio  $Z^S_m/Z^{NS}_m$
 is to go back to one of the pioneering dynamical
 calculations~\cite{Eicker97}, 
 carried out with two flavours of unimproved Wilson fermions. 
 In this paper they initially found a value for the ratio
 $m_s/m_l \approx 52$, which is a long way from the value
 found in quenched calculations, and also far from the predictions
 of chiral perturbation theory
 \begin{equation}
 \frac{m_s}{m_l} = \frac{ 2 m_K^2 - m_\pi^2}{m_\pi^2} \approx 25\;.
 \label{cpratio} 
 \end{equation}  
 Normally one would expect the mass renormalisation factor  to cancel
 out in this ratio, but because the singlet and non-singlet $Z$
 differ this is not entirely true. 
 
     In Fig.~{\ref{slightratio}} we show a pion line, the line  
 giving the pseudoscalar mass when valence and sea quark masses are equal, 
 and the kaon line, showing the result when sea quarks and one valence quark
 (representing the $u$ or $d$ in the $K$)   
 are kept at a fixed mass, while the other valence quark (representing 
 the strange quark) is varied. The slopes of these two lines are 
 different, just as seen in the partially quenched case. 
   
 \begin{figure}[htbp]
 \begin{center}
 \includegraphics*[width=0.30\textwidth,angle=270]{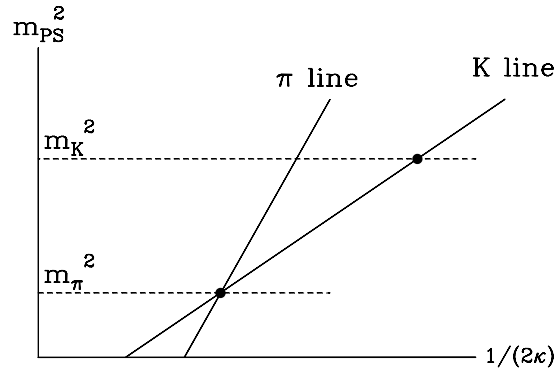}
 \end{center} 
 \caption{\it A schematic diagram illustrating the subtleties
 of measuring the strange/light quark mass ratio from Wilson
 fermions. 
 \label{slightratio} }
 \end{figure} 

   We determine the $\kappa$ corresponding to the light quark mass 
 by seeing where the $\pi$ line crosses the mass-squared of the 
 real physical pion, and the strange quark $\kappa$ by seeing where
 the kaon line crosses the mass-squared of the physical $K$. 
 If we now use these two $\kappa$ values to calculate the ratio
 of the {\it bare} quark masses, $m^{bare}_s/m^{bare}_l$ we get 
 a ratio much larger than from eq.~(\ref{cpratio}), because the $K$ line 
 in Fig.~\ref{slightratio} is  less steep than the $\pi$ line. 
 However, if we renormalise the masses according to
 eq.~(\ref{renrecipe}), then the slopes of the kaon and pion 
 lines will become approximately equal, as in the lower panel
 of Fig.~\ref{FullPartial}, and now the ratio of the 
 {\it renormalised} quark masses will be close to eq.~(\ref{cpratio}). 
   
 \begin{figure}[htbp]
 \begin{center}
 \includegraphics*[width=0.48\textwidth]
 {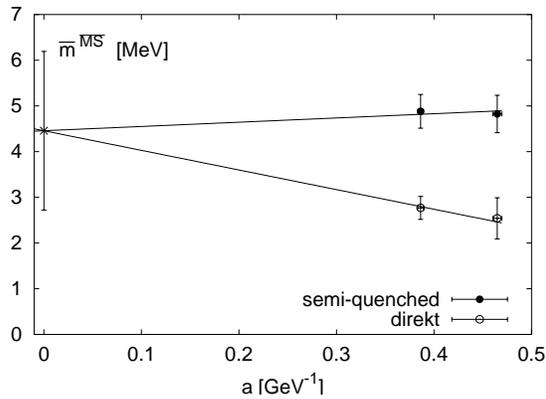}
 \end{center} 
 \caption{\it A figure from~\cite{SESAM2} showing the difference
 in light quark masses found from partially quenched and
 fully dynamical  (``direkt'') data. 
 \label{PQdirect} }
 \end{figure} 
 
 In later works~\cite{SESAM2} the SESAM collaboration used a
 different 
 strategy, measuring data from both partially quenched and fully
 dynamical mesons, and then making a joint extrapolation to the 
 continuum limit, Fig.~\ref{PQdirect}. The same $Z$ was used in both cases.
 This gives a result
 more in line with what we do here, but it is not exactly the same. 
 Although we have seen in Fig.~\ref{rationew} that the ratio
 $Z^S_m/Z^{NS}_m$ decreases as $a$ decreases, there will 
 presumably be a perturbative contribution at the two-loop level, 
 which means that the $Z$ ratio will asymptotically approach the
 value 1 rather slowly, like $1 + O(g^4)$, rather than like a power
 of $a$, so one will not be able to completely remove this effect
 by power-law extrapolations in $a$. 

    The argument we give here, using different $Z$ factors
 for the singlet and non-singlet parts of the mass matrix
 is rather similar to the discussion given at Lattice~'97
 in~\cite{Bhattacharya}.

 \subsection{Quark mass results}

   The preliminaries are now over, and it is time to discuss
 some of the recent quark mass results from dynamical simulations. 

 \begin{figure}[htbp]
 \vspace*{-1.3cm}
 \begin{center}
 \includegraphics*[width=0.5\textwidth]{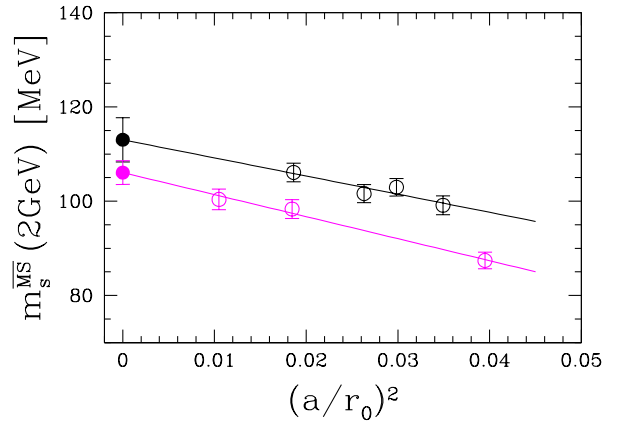}
 \vspace*{-9mm}
 \end{center} 
 \caption{\it A comparison of quenched (above)
 and dynamical (below)
 estimates for the strange quark mass~\cite{QCDSFmq}.
 \label{qmassescom} }
 \end{figure} 

   In Fig.~{\ref{qmassescom}} we show an estimate of the strange
 quark mass calculated using the flavour-singlet renormalisation 
 constant calculated from the comparison of dynamical and 
 partially quenched mesons. The result is rather similar to
 that found in the quenched case. If the non-singlet renormalisation 
 constant had been used the $a$-dependence would be much stronger, 
 and the data points would have been lower. 

   An alternative mass definition is based on the 
 continuum Ward identity for an axial current with flavour
 index $a$~\cite{MM}, 
% \clearpage
 \begin{equation}
 \langle \partial_\mu A_\mu^a \rangle 
 = \langle \psib 
 \left\{ m , \tau^a \right\} \gamma_5 \psi \rangle + 
 \hbox{contact terms}. 
 \end{equation} 
 In clover fermions the axial current has to be improved
 by adding irrelevant operators, because of the lack of 
 a true chiral symmetry. 
 As can be seen from the Ward identity, this axial Ward identity (AWI)
 mass  has to be renormalised by the factor
 \begin{equation}
 m^{ren} = \frac{Z_A}{Z_P} m^{AWI} \;. 
 \end{equation}
 The final renormalised result ought to be the same in either case, 
 but we will see that this is not yet the case, there are still 
 fairly large differences between different groups working on this
 problem. 

\begin{figure}[htbp]
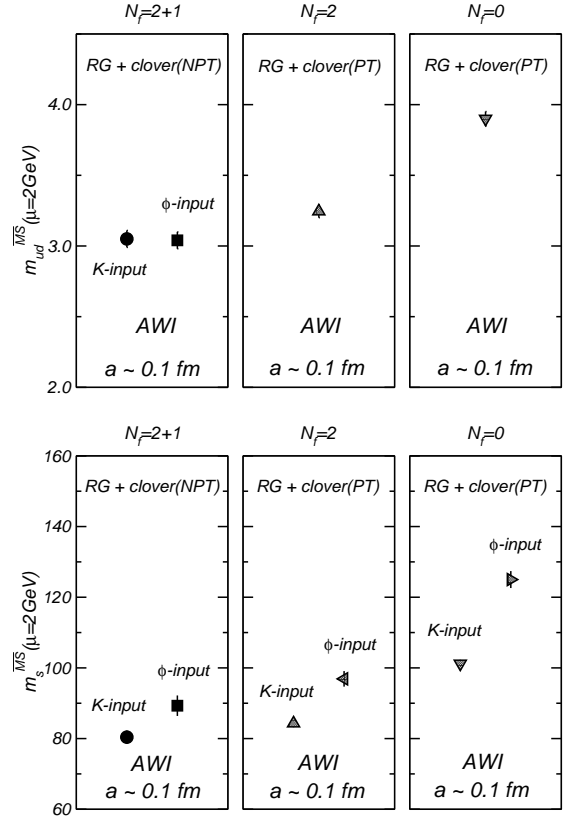

 \begin{center}
  \includegraphics*[scale=0.32]
  {Mud_vs_Nf.eps}
 \end{center}
 \begin{center}
  \includegraphics*[scale=0.32]
  {Ms_vs_Nf.eps}
 \end{center}
 \vspace*{-10mm}
 \caption{\it The $u$ and $d$ quark masses (above) and $s$ quark mass (below)
 at $a\sim 0.1$~fm, from~\cite{Ishikawa_mq}.}
 \label{CPPACS_JLQCD}
 \vspace*{-3mm}
\end{figure}

  The AWI quark mass definition has been used by the
  CP-PACS and JLQCD  Collaborations to study the quark masses
 in clover QCD, both for two-flavour~\cite{AliKhan} and (2+1) flavour 
 simulations~\cite{Ishikawa_mq}. The $Z$ ratio needed to renormalise
 the bare lattice masses are calculated from tadpole-improved
 one-loop lattice perturbation theory. 
 Their results are summarised in Fig.~\ref{CPPACS_JLQCD}. 
 In the quenched case the result depends quite heavily on whether
 the $K$ or $\phi$ meson is used to determine the strange quark mass. 
 This is a sign that the quenched theory differs from the real world. 
 Fortunately this discrepancy is greatly reduced when sea quarks are
 switched on, as can be seen in the panels of  Fig.~\ref{CPPACS_JLQCD}
 showing the $s$ quark mass from dynamical QCD. 

   The SPQcdR Collaboration~\cite{Rome_mq} have measured quark masses in an
 $N_f=2$ simulation using the Wilson quark action at a single
 $\beta$ value, $\beta=5.8$. They use both the VWI and AWI 
 definitions of quark mass, and find compatible results from
 both methods. The renormalisation constants are found non-perturbatively
 using the method of~\cite{Martinelli:1994}. Their preliminary
 values are included in Table~\ref{mls}, and are broadly compatible
 with the results of~\cite{QCDSFmq}.

   In~\cite{Stagger_mq}  
 the quark masses are measured in a dynamical simulation
 with (2+1) flavours of sea quark. The use of staggered 
 quarks speeds up the calculation considerably, allowing the 
 light $u/d$ sea quarks to be simulated all the way down to
 masses $\sim m_s/8$, a considerable advance on what is possible with
 other quark formulations. The staggered quarks have been improved
 to suppress `taste' violation and other discretisation errors. 
 This also has the effect of making  perturbation theory better 
 behaved than it is for unimproved staggered fermions. 
 The $Z$ factors used in~\cite{Stagger_mq} are calculated in 
 one-loop tadpole improved perturbation theory. The estimated 
 error due to the unknown higher-order terms in perturbation 
 theory is 9\%.  This study finds that the effects of including
 sea quarks are dramatic, and that the final physical
 masses of the quarks are a lot lighter in the dynamical simulation
 than they were in a quenched calculation.

 \begin{table*}[htbp]
 \begin{center}
 \begin{tabular}{|c|c|c|c|} 
 \hline
 $m_l$/MeV & $m_s$/MeV  & sea quarks & Reference \\ 
 \hline
 3.05(6) &80.4(1.9) & (2+1) & CP-PACS and JLQCD \cite{Ishikawa_mq} \\
 2.8(0)(1)(3)(0) &76(0)(3)(7)(0)  & (2+1) & 
 HPQCD, MILC and UKQCD \cite{Stagger_mq}\\
  4.7(2)(3)  &119(5)(8) & $N_f=2$ & QCDSF/UKQCD \cite{QCDSFmq} \\
 4.8(5) & 111(6) &  $N_f=2$ &  SPQcdR Collaboration  \cite{Rome_mq} \\
 4.5(17) &        &  $N_f=2$ & SESAM/T$\chi$L \cite{SESAM2} \\
 \hline
 \end{tabular} 
 \end{center} 
 \caption{ \it Masses of the light, $m_l = \frac{1}{2}(m_u + m_d)$,
 and strange quarks
 in dynamical lattice QCD, as determined by various collaborations. 
 The masses are given in the $\msb$ scheme at a scale of 2 GeV. 
 \label{mls} } 
 \end{table*}

 \begin{figure}
 \begin{center}{ 
 \includegraphics*[width=7.0cm,clip=]{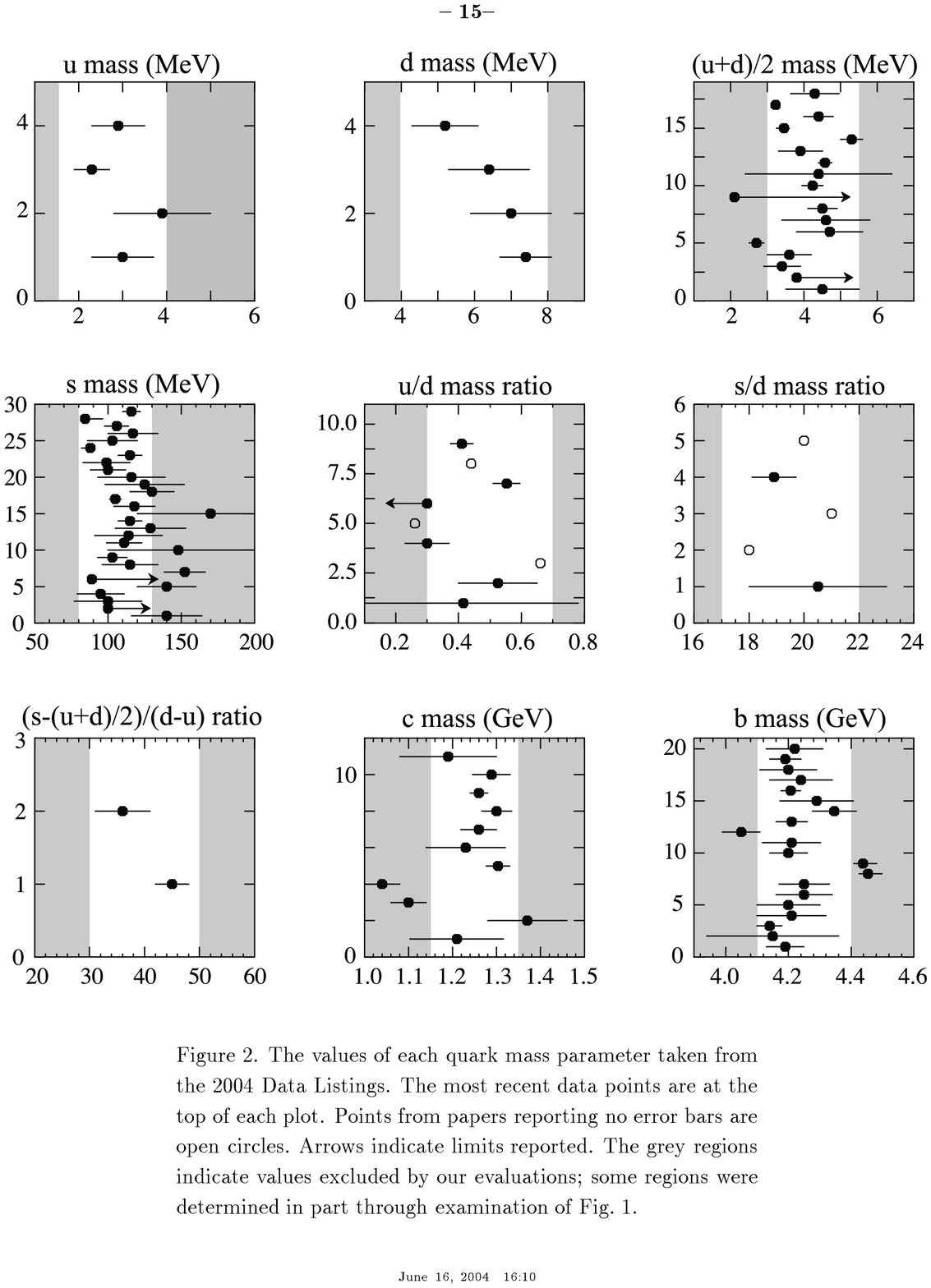} }
 \end{center}
 \caption{\it Determinations of the strange quark mass 
 from the Particle Data Group~\cite{PDG}. \label{pdgcomp}} 
 \end{figure} 

  Fig.~\ref{pdgcomp} shows determinations of the strange quark mass
 compiled by the Particle Data Group~\cite{PDG}. Comparing this with
 Table~\ref{mls} we see that most of the dynamical lattice determinations
 look low compared with estimates from traditional methods. 

 \subsection{The charm quark mass}

    The charm quark mass has its own special difficulties
 because $m_c$ is comparable with the inverse lattice spacing
 $a^{-1}$ for the lattice spacings we have to work with at present. 
 This means that lattice artefacts proportional to high powers
 of $a m_c$ can be important, so it would be useful to have 
 expressions resumming all $(a m_c)^n$  terms. Suggestions on
 how to do this (known as FNAL masses) were made in~\cite{FNALmass}. 
     As the lattice spacing becomes smaller all definitions 
 should agree. This is examined in~\cite{Maynard}, where the charm
 mass determined from the axial and vector Ward identity
 definitions, 
 and the mass determined from the FNAL method $m_1$ and $m_2$
 are compared in the quenched case (where the $a$ range can be 
 made large). 
 At present lattice spacings the mass from the 
 Axial Ward Identity ($m_A$) looks as if it might extrapolate to 
 a lower value than the others, though there is still room for it 
 to move up again, see Fig.~\ref{Maynard}. 

 \begin{figure}[htbp]
 \begin{center}
    \includegraphics*[width=7cm,clip]{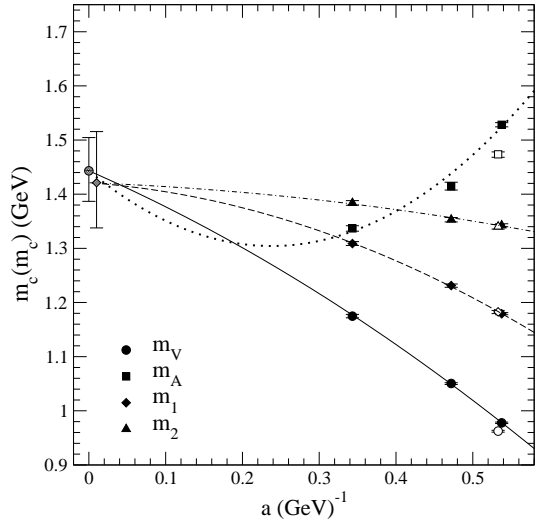}
 \end{center} \caption
 {\it Different determinations of the charm quark mass~\cite{Maynard}. 
 The closed symbols show the effect of using different mass definitions
 in the quenched case. All definitions should lead to the same 
 continuum value. The open symbols at the right show the charmed 
 quark mass determined on $N_f=2$ dynamical configurations from UKQCD.
 \label{Maynard} }
 \end{figure}

   This reference also looks at the effect of finding $m_c$
 on dynamical clover configurations, the open symbols on the 
 right-hand side of Fig.~\ref{Maynard}. There seems to be very little
 difference between the quenched and dynamical results, though of course
 the present results are at a single rather coarse lattice spacing, 
 and the sea quarks in the simulation are rather heavy. 

     \subsection{$b$ quark mass}

  A nice example of the interplay  between perturbation theory 
 and physical result is the $b$ quark mass calculation of 
 Gimenez, Giusti, Rapuano and Martinelli~\cite{bGimenez}
 converting lattice data into a value for $m_b$ in $\msb$, 
 using stochastic perturbation theory results from~\cite{bDiRenzo}. 

    This requires perturbation theory in two places, the conversion
 from the lattice mass parameter of heavy quark effective theory
 (HQET) to the pole mass requires the subtraction of a residual
 mass $\delta m$ which is proportional to $1/a$. This subtraction 
 can be calculated in lattice perturbation theory. After that, 
 the conversion from pole mass to $\msb$ mass is done using results
 from continuum perturbation theory. Since the $b$ quark pole mass 
 is just used as an intermediate step the calculation taken as a whole
 is presumably legitimate even though confinement means that 
 there is no pole in the $b$-quark propagator.  

    The two-loop subtraction for the residual mass was calculated 
 using traditional lattice perturbation theory  methods
 in~\cite{db2loop}. The three-loop result was calculated using
 the novel method of stochastic perturbation theory, which 
 has the major advantage that the difficulty of calculating new 
 terms in the series does not increase as rapidly as it does in
 conventional calculations. Stochastic perturbation theory thus 
 promises to overtake conventional techniques in the future. 
 It does have some disadvantages, being a stochastic method, the 
 coefficients calculated do have a statistical error, and can
 not be given to the number of decimal places that we are used to
 from conventional perturbation theory, but they are usually
 quite good enough for comparison with real lattice data. 
 
    The dependence of the final $b$ quark mass on the order of
 lattice perturbation theory used is illustrated in Table~\ref{btable}. 
 Knowing the three-loop coefficient helps to significantly reduce the
 systematic uncertainty in the final estimate, which is 
 $$ \overline{m}_{b}(\overline{m}_{b})|_{unq}\, =\, 
 (4.21\, \pm\, 0.03\, \pm\, 0.05\, \pm\, 0.04)\;\mbox{\rm GeV}. $$ 

\begin{table*}
{\centering \begin{tabular}{|c||c|c||c||c|c|}
\hline
\( \overline{m}_{b}(\overline{m}_{b}) \)&
\multicolumn{5}{|c|}{{Dependence on higher orders (unquenched data) }
 }\\
\hline
\hline
{Order}&
\( \delta m \)&
%\( m^{pole}_{b}=M_{B_{s}}-{\cal E}_{B_{s}}+\delta m \)&
 \( m^{pole}_{b} \)&
\( \overline{m}_{b}(\overline{m}_{b})/m^{pole}_{b} \)&
\( \overline{m}_{b}(\overline{m}_{b}) \)&\(\Delta m_{b}\)\\
\hline
LO&
\( 0 \)&
\( 3770 \)&
\( 1.000 \)&
\( 3770 \)&
--\\
\hline
NLO&
\( 978.79 \)&
\( 4748.79 \)&
\( 0.92276 \)&
\( 4382 \)&
\(612\)\\
\hline
NNLO&
\( 10.53 \)&
\( 4759.32 \)&
\( 0.89060 \)&
\( 4239 \)&
\(143\)\\
\hline
NNNLO&
\( 74.55 \)&
\( 4833.87 \)&
\( 0.86883 \)&
\( 4200 \)&
\(39\)\\
\hline
\end{tabular}\par} % \vspace*{0.8cm}
 \caption{ \it A table illustrating the way in which the
 estimate for the bottom quark mass $m_b(m_b)$ depends on the
 order of the lattice perturbation theory used.
 The quark mass results are preliminary values
  from~\cite{bGimenez}, using
 perturbative calculations from~\cite{bDiRenzo} and ~\cite{db2loop}. 
 \label{btable} } 
 \end{table*}

%-------------------------------------------------

  \section{DETERMINING  $\Lambda$}

 An important (but difficult) project for the lattice is
 to link the low-energy, non-perturbative quantities we
 normally measure (such as hadron masses) with the
 high-energy world accessible in accelerators, where 
 perturbative QCD works well. The key to bridging this
 gap is to find a value for $\Lambda$, the scale parameter 
 of QCD. 

   The strategy for determining $\Lambda$ has 
  two necessary parts. 
  Firstly some dimensionful  quantity has to be measured
 to set the scale. A common choice is to look at the 
 static potential. This can be measured fairly well. 
 A good way to fix a length scale is to find the distance
 at which
 \begin{equation}
  r^2 F(r) \equiv -r^2 \frac{d}{d r} V(r) 
 \end{equation} 
 has a particular value~\cite{r0_Sommer}, the usual value 
 chosen is 1.65, which gives a distance scale $r_0 \approx 0.5fm$. 
 Another way to find a length scale from the potential,
 now less popular, would be to use the string tension. 

    Secondly, we need to extract the $\msb$ coupling at a known 
 scale, which can then be converted into a value for  $\Lambda_{\msb}$. 
  One possibility is to use lattice perturbation theory to
 relate the lattice coupling  $g^2 = 6/\beta$ to  $g^2_{\msb}$.
 
 Relating the $\Lambda$ parameters of two schemes 
 is a finite calculation. This is an asymptotic statement, 
 and can be answered completely by a one-loop
 calculation~\cite{Lambdaratio}.

 However, relating the coupling constants at a finite scale
 is no longer so simple. 
 Outside the asymptotic regime we need calculations of
 the $\beta$-function
 of both schemes to as high an order as possible, this can then
 be translated into a relation between the
 couplings. Currently we know the lattice $\beta$ function
 to three loops~\cite{PanagopoulosBeta}.
 
     As often happens with lattice perturbation theory 
 we find that the series converge rather poorly.  Fortunately
 we have several methods to accelerate the convergence. 
 One explanation for the large coefficients in lattice perturbation
 theory is that they come from gluon ``tadpoles'' due to the 
 higher order vertices present on the lattice~\cite{Lepage}.
  A way to take these
 into account is to re-express the series in terms of a ``boosted''
 coupling,  $ g^2_{\Box} = g^2/ U^{plaq} $: 
 \begin{eqnarray}
  \frac{1}{g^2_\msb(\frac{1}{a})} &=& \frac{ 1}{g^2}
  -0.4682013 -0.0556675 g^2 ,\nonumber \\
    \frac{ 1}{g^2_\msb(\frac{1}{a})} & =& \frac{ 1}{g^2_\Box}
  - 0.1348680 -0.0217565 g^2_\Box ,\nonumber \\ 
  \frac{ 1}{g^2_\msb( \frac{2.63}{a})} &= &\frac{1}{g^2_\Box} \qquad 
 \quad -0.013837  g^2_\Box .
 \label{imp_ladder} 
 \end{eqnarray} 
 Using the quenched case as an example 
 we can see in the equation above that the coefficients in the 
 series for the $\msb$ coupling are reduced when we change 
 the expansion parameter from $g^2$ to $g^2_{\Box} $. 
 We can improve the series further by choosing a natural 
 scale~\cite{Llect}. If, instead of $g^2_\msb(1/a)$ we calculate 
 $g^2_\msb(2.63/a)$, the $g^0_\Box$ term vanishes, and the first
 correction is $O(g^2_\Box)$ with a rather small coefficient.  
 This gives us grounds to hope that the unknown higher-order 
 coefficients will not seriously change the estimate of $g^2_\msb$. 
 
  Once we have values for  $g^2_\msb$ we can use the known
 $\beta$ function to convert these into a value for $\Lambda^{\msb}$.
 The results in the quenched case~\cite{Lambda_update},
  based on data from~\cite{BoothLambda}
  and~\cite{NeccoSommer}, are shown in Fig.~\ref{QuenchLambda}.   

  \begin{figure}[htbp]
 \vspace*{-5mm}
 \includegraphics*[width=0.45\textwidth]{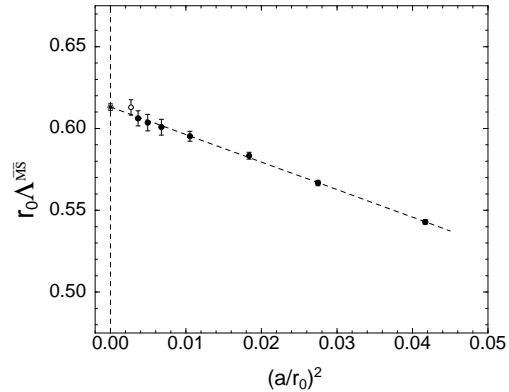}  
 \vspace*{-9mm}
 \caption{ \it $\Lambda^{\msb}$ calculated for quenched QCD
 for $\beta$ values
 in the range 5.95 to 6.92~\cite{Lambda_update}. 
 The $r_0$ values for high $\beta$ are from~\cite{NeccoSommer}. 
 The data extrapolates very  smoothly towards  a value 
   $\Lambda^\msb_{N_f=0}=242(11)(10) MeV$.  
 \label{QuenchLambda} }
 \vspace*{-4mm}
 \end{figure} 

    Some extrapolation in $a$ is needed, this seems not to
 present any great problem, the data look surprisingly
 linear in $a^2$, and the smallest $a$ values are 
 close to the continuum limit. The result in physical
 units is  $\Lambda^\msb_{N_f=0}=242(11)(10) MeV$.
  The first error is a statistical error, the second 
 error is an attempt to estimate the possible effects of
 the unknown higher order terms in the $\beta$ functions. 

   One can try a similar calculation for the dynamical case,
 using the clover action configurations produced by the 
 UKQCD/QCDSF Collaborations. 
 The lattice $\beta$ function for clover quarks is
 available~\cite{PanagopoulosBeta}. In the dynamical 
 case we need a double extrapolation, both in lattice
 spacing and in sea quark mass. Both these extrapolations
 are over a rather large distance.  A simple phenomenological
 extrapolation formula is 
 \begin{equation} 
  r_0 \Lambda^\msb = const + B (a/r_0)^2 + C a m_q + D r_0 m_q \;. 
 \end{equation} 
 The formula fits the data fairly well, as can be seen in 
 Fig.~\ref{DynLambda}. 

  \begin{figure}[htbp]
 \begin{center}
\includegraphics[width=0.5\textwidth] 
 {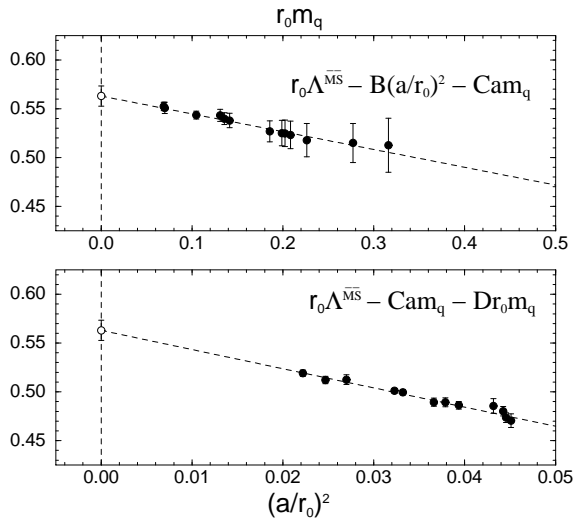}
 \end{center} 
 \vspace*{-9mm}
 \caption{ \it Extrapolation of $\Lambda^\msb$ for dynamical
 quarks. The data are measured on UKQCD/QCDSF configurations
 with dynamical clover fermions~\cite{Lambda_update}.  
 The upper part of the figure shows the mass dependence
 after the data has been extrapolated to the continuum limit. 
 The lower half shows the $a$ dependence of data after extrapolating
 to the chiral limit. 
 \label{DynLambda}} 
 \end{figure} 
  
%  $r_0 \Lambda$ much more dependent on sea-quark mass than 
% perturbation theory expected. 

    The extrapolated value hasn't changed very much due to the jump
 to finer lattice spacings in the past year~\cite{Lambda_update},
 the new value is   
 $\Lambda = 222(4)(28) MeV$ at $N_f=2$. 

 \begin{figure}[htbp] 
 \begin{center}
 \includegraphics[width=7cm]{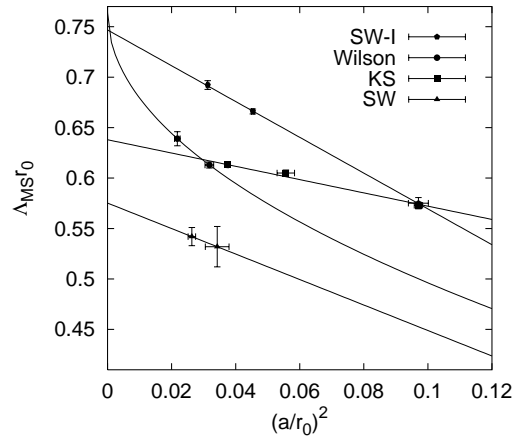}
 \end{center}
 \vspace*{-9mm}
 \caption{\it A comparison of continuum extrapolations
 of $r_0 \Lambda^\msb$~\cite{BaliBoyle} made for various fermion actions.
 Wilson fermions are extrapolated linearly in $a$, 
 while clover and staggered are extrapolated linearly in
 $a^2$.\label{BBextrap}} 
 \vspace*{-4mm}
 \end{figure} 

    One way to estimate the systematic error in the final result 
 is to repeat the calculation for different formulations of dynamical
 fermions, and see how much the results scatter. A very complete study was 
 carried out in~\cite{BaliBoyle}. As can be seen in Fig.~\ref{BBextrap}
 the results do still depend quite a lot on the action and extrapolation
 procedure. The dependence on the action could be due in part to higher
 order terms in the perturbation series, which will of course depend
 on the fermions used. 
 
   A very impressive attack on finding $\Lambda$ was reported in 
 this year's Lattice Conference~\cite{Mason04},
 extending work done in~\cite{HPUKlat02} with 
 Asqtad improved staggered fermions. Extensive three-loop
 calculations of Wilson loops and the static potential have been
 carried out. Some idea of the work involved can be gained by looking at 
 the number of Feynman diagrams that have to be considered, 
 Fig.~\ref{feyn}. Wilson loops of different sizes are dominated
 by gluons of different virtualities, with small loops corresponding
 to large $Q^2$. Thus by calculating $\alpha_s$ from loops of 
 different sizes one can see how $\alpha_s$ runs even from data
 taken at a single value of $a$. This is illustrated in 
 Fig.~\ref{masonrun}, where one can see how well the running of
 $\alpha$ agrees with the $\beta$ function predictions. 

 \begin{figure}[htbp] 
 \begin{center}
 \vspace*{4mm}
\includegraphics[width=0.45\textwidth,clip]{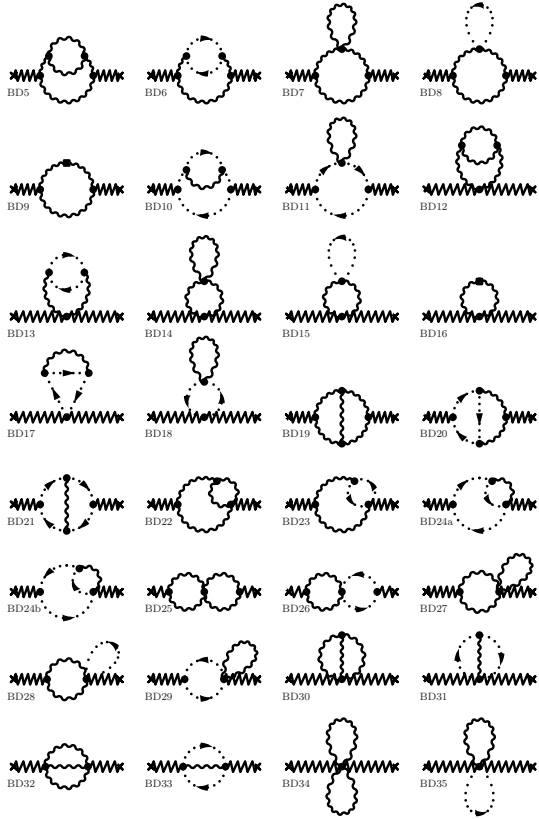}
  \caption{ \it The gluon and ghost 
 Feynman diagrams used in~\cite{Mason04} to calculate 
 three-loop results for Wilson loops and the static potential.
 A similar number of fermion loop diagrams is needed to calculate
 the unquenched case. \label{feyn}} 
 \end{center}
 \end{figure}

 \begin{figure}[htbp]
 \begin{center}
\includegraphics[width=0.33\textwidth, angle=270]{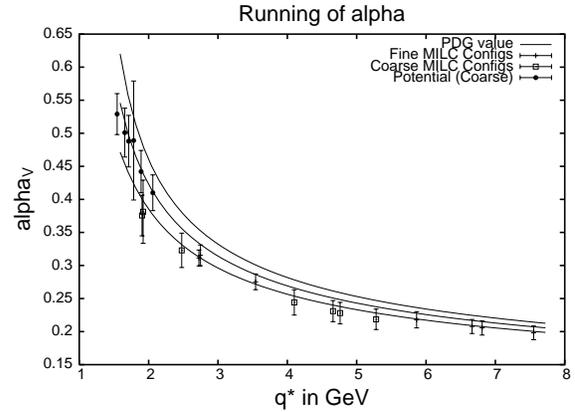}
 \caption{ \it The running of $\alpha_s$ as determined by~\cite{Mason04}.
 $\alpha_s$ is defined in the potential scheme.
 \label{masonrun} } 
 \end{center}
 \end{figure}
 
 In Fig.~\ref{Mason_amZ} we see the lattice results extrapolated 
 out to the scale $m_Z$, where they are compared with the traditional 
 Particle Data Group value. The agreement looks impressively good. 

%-------------------------------------------------
   
  \begin{figure}[htbp] 
 \begin{center}
\hspace*{-5mm}\epsfig{file = 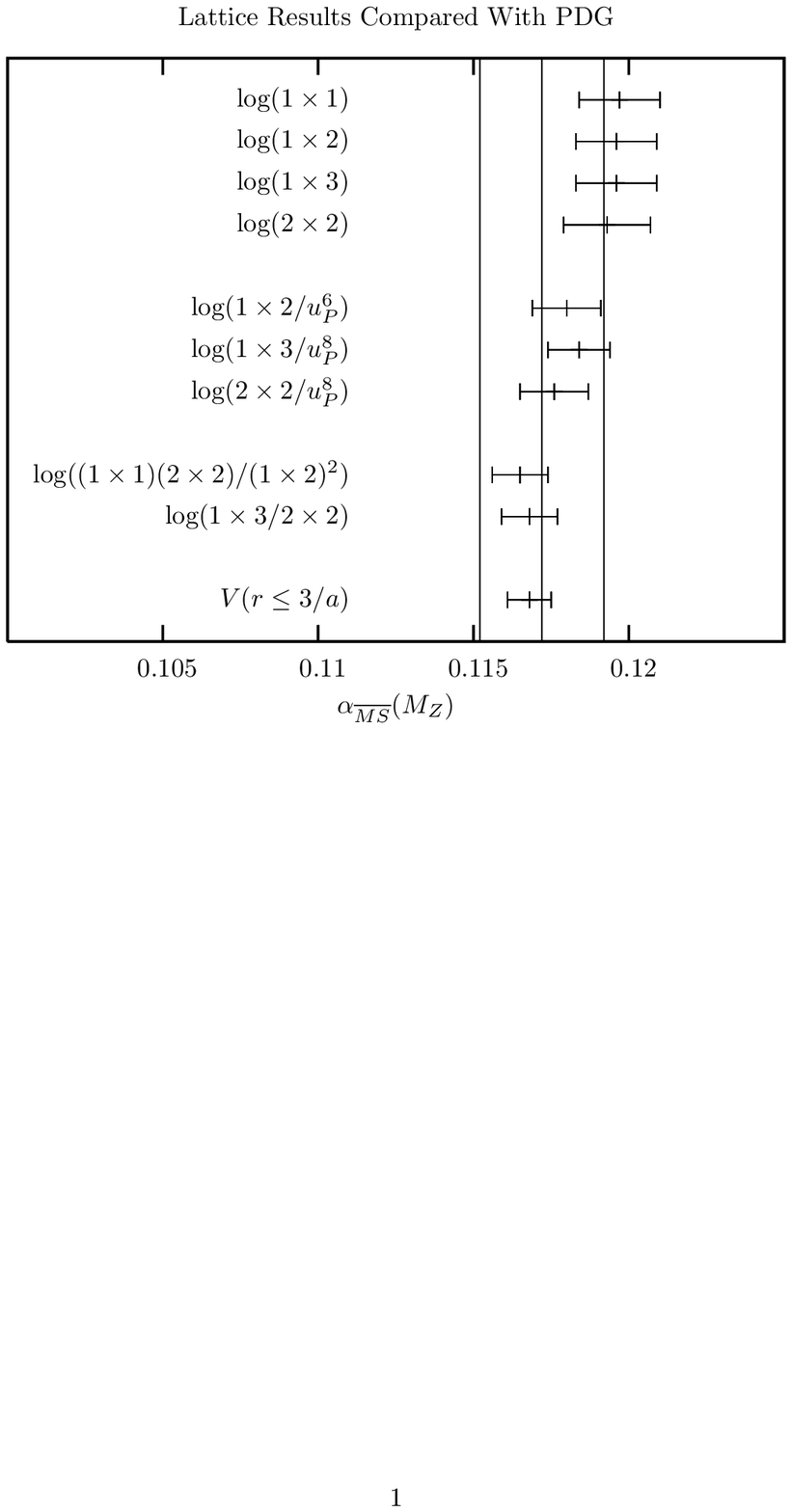, clip=, width= 8.3cm}
 \caption{ \it $\alpha_s(m_Z)$ estimated from Wilson loops
 of various sizes, and from the short distance potential~\cite{Mason04}.
 \label{Mason_amZ}}
 \end{center}
 \end{figure}

%-------------------------------------------------

%
% \subsection{ Finding Dynamical  $\Lambda$ from the 
%     Schr\"odinger functional} 
%

   Is there any way of diminishing the dependence on high-order 
 perturbation theory? The obvious way to reduce our sensitivity
 to high order terms in the perturbation theory is to work at 
 weaker values of the coupling --- in QCD this means working at 
 high energy or short distance.  This requires a very fine
 lattice spacing, $a$, which in turn forces us to consider
 simulations in small physical volumes. The small volumes 
 involved in this method mean that it isn't possible to reuse
 configurations generated for other projects on hadronic physics,
 the calculations require many dedicated runs at very 
 large $\beta$ values. 

  \begin{figure}[htbp]
 \vspace*{-8mm}
 \begin{center}
 \hspace*{-1cm}\epsfig{file=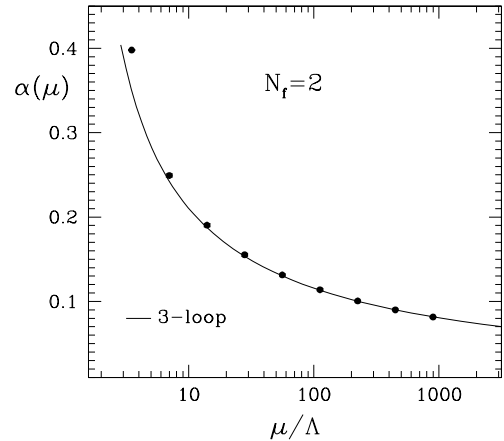, width=9cm}
 \end{center} 
 \vspace*{-8mm}
 \caption{\it The running of $\alpha_s$ in two-flavour dynamical QCD
 as determined by the ALPHA Collaboration~\cite{alphaLambda}.
 \label{run_alpha} } 
 \end{figure} 

 The ALPHA Collaboration have been pursuing this method for 
 many years~\cite{alphaLambda}, using the same methods that
 they applied in the quenched case~\cite{CapitaniALPHA}. 
 A recent plot of their results is shown in Fig.~{\ref{run_alpha}}. 
 By repeatedly halving the lattice spacing  (and the 
 physical size $L$ of their system), they make
 measurements of $\alpha_s$ over a large range of scales (more than
 two orders-of-magnitude). These are compared with the running expected 
 from the three-loop $\beta$ function. The data reach $\alpha_s$
 values $\sim 0.1$, where we are very confident that perturbation
 theory will work well. As can be seen, the agreement with
 the perturbative $\beta$ function remains close, except perhaps at the 
 very lowest scale. The fact that perturbation theory works so well
 over such a large range 
 offers some comfort to the groups that base their determination of
 $\Lambda$ on lattices with ``normal'' lattice spacings, 
 who have to rely on data with $\mu/\Lambda$ around 10 or 20,
 but clearly the precision that will be achieved from data at 
  $\mu/\Lambda \sim 1000$ should be much greater.  

    This work has not yet given a final value for $\Lambda$ in
 physical units, preliminary results give the value
  $\ln (\Lambda L_{max}) = -1.34(7)$ where 
  $\alpha_s(L_{max}) = 0.372$, but the final step is still 
 needed, a conversion from the 
 length scale they use, $L_{max}$, to  physical units.

%-------------------------------------------------

   \section{ CONCLUSIONS}
 
   The determination of quark masses is the more mature of
 the two fields we have looked at.  We see that agreement
 has not yet been reached, and there are still differences
 between the measurements of the different groups. It is 
 not yet clear (to me at least) where the origin of these
 differences lies, as there are many differences between the 
 calculations, particularly in the ways the $Z$ factors are 
 calculated and in the choice of mass definition. 

    One area in which we are making progress, and can expect more
 advances in the coming years, is bringing the sea quark masses 
 down towards more realistic values. When simulations are carried
 out with large sea quark masses it can be difficult to find much
 difference from quenched calculations, and the extrapolation to 
 the physical region can be difficult. Now, with more powerful 
 machines and new fermion actions, we are seeing simulations 
 carried out much closer to the physical region, and we can expect more 
 soon. 
   
   A topic which we have not covered here is the calculation
 of quark masses using overlap fermions, since dynamical overlap
 is not yet as advanced as dynamical calculations with the older fermion
 formulations. 

   In both the calculation of masses and of $\Lambda$ we have a 
 need for higher orders of perturbation theory.  Advances in 
 lattice simulations are making lattice perturbation theory more
 difficult, because more complicated actions lead to more 
 complicated Feynman rules, and because there is now far greater 
 diversity in the choice of fermion and gauge actions used.   
 At Lattice~'04 we have seen that diagrammatic techniques are
 rising to the challenge, we have also seen that 
  stochastic perturbation theory is delivering results which
 are useful to those extracting physics from lattice simulations.

%  \section{ Acknowledgements}
%

\end{document}